\pdfoutput=1
\documentclass[conference]{IEEEtran}
\usepackage[T1]{fontenc}
\usepackage{amsmath}
\usepackage{amsfonts}
\usepackage{amssymb}
\usepackage{cite}
\usepackage{multicol}
\usepackage{verbatim}
\usepackage{graphicx}
\usepackage{subcaption}
\usepackage{balance}
\usepackage{epstopdf}
\usepackage{epsfig,color}
\usepackage{float}
\usepackage{subcaption}
\usepackage[linesnumbered,ruled]{algorithm2e}

\SetKwInOut{Parameter}{Parameters}
\SetKwRepeat{Do}{do}{while}

\IEEEoverridecommandlockouts
\begin{document}

%\title{Resource optimization (analytical + learning based) for secrecy rate maximization in underwater acoustic sensor networks}

%\title{Node selection and power allocation for secrecy rate maximization in OFDM-based multihop underwater acoustic sensor networks}

\title{Securing the Insecure: A First-Line-of-Defense for Nanoscale Communication Systems Operating in THz Band}
\author{
\IEEEauthorblockN{ Waqas Aman$^\ast \dagger$, %\IEEEauthorrefmark{1}
M. Mahboob Ur Rahman$^\ast$%\IEEEauthorrefmark{1}
, Hassan\ T.\ Abbas$^\dagger$%\IEEEauthorrefmark{1}
, Muhammad Arslan Khalid$^\bot$, Muhammad\ A.\ Imran$^\dagger$, \\ Akram\ Alomainy$^{\dagger \dagger}$, and Qammer\ H.\ Abbasi$^\dagger$ %\IEEEauthorrefmark{1} 
} 
\IEEEauthorblockA{%\IEEEauthorrefmark{1}
$\ast$ Electrical
Engineering Department, Information Technology University, Lahore, 54000,
Pakistan\\ 
$\dagger$Department of
Electronics and Nano Engineering, University of Glasgow, Glasgow, G12 8QQ,
UK \\
$\bot$Division of Biomedical Engineering, School of Engineering, University of Glasgow, Glasgow, G12 8QQ,
UK\\ 
$\dagger \dagger$School of Electronic Engineering and Computer science, Queen Mary University of London, London, E1 4NS, UK
\\
$^\ast$\{waqas.aman, mahboob.rahman\}@itu.edu.pk, $^\dagger$\{Hassan.Abbas,  Muhammas.Imran, Qammer. Abbasi\}@glasgow.ac.uk,\\ $^\bot$arslan.k@live.com, $^{\dagger \dagger}$a.alomainy@qmul.ac.uk }
}

\maketitle % make the title area

%\section{Abstract}

\begin{abstract}
Nanoscale communication systems operating in Terahertz (THz) band are anticipated to revolutionise the healthcare systems of the future. Global wireless data traffic is undergoing a rapid growth. However, wireless systems, due to their broadcasting nature, are vulnerable to  malicious security breaches. In addition, advances in quantum computing poses a risk to existing crypto-based information security. It is of the utmost importance to make the THz systems resilient to potential active and passive attacks which may lead to devastating consequences, especially when handling sensitive patient data in healthcare systems. New strategies are needed to analyse these malicious attacks and to propose viable countermeasures. In this manuscript, we present a new authentication mechanism for nanoscale communication systems operating in THz band at the physical layer. We assessed an impersonation attack on a THz system. We propose using path loss as a fingerprint to conduct authentication via two-step hypothesis testing for a transmission device. We used hidden Markov Model (HMM) viterbi algorithm to enhance the output of hypothesis testing. We also conducted transmitter identification using maximum likelihood and Gaussian mixture model (GMM) expectation maximization algorithms. Our simulations showed that the error probabilities are a decreasing functions of SNR. At $10$ dB with $0.2$ false alarm, the detection probability was almost one. We further observed that HMM out-performs hypothesis testing at low SNR regime ($10 \% $ increase in accuracy is recorded at SNR = $−5$ dB) whereas the GMM is useful when ground truths are noisy. Our work addresses major security gaps faced by communication system either through malicious breaches or quantum computing, enabling new applications of nanoscale systems for Industry 4.0. 
%The simulation results show that: i) the proposed schemes outperform the other schemes (i.e., the performance gap between the proposed and other schemes increases) with increase in the transmit power budget of the sensor nodes, ii) the proposed schemes benefit from the increase in the density of sensor nodes (while the depth base selection  scheme fails in such situation).

\end{abstract}

\section{Introduction}

The recent developments in the nano fabrication technologies has led to an increased interest in the design of nanoscale communication systems where small devices (of size few nm) that are few millimeter apart from each other communicate to each other \cite{Islam:CM:2010}. Due to their small size, the existing frameworks, techniques and methods proposed for communication networks such as Wifi, 4G etc. are not suitable for exchanging information amongst the nano devices \cite{AKYILDIZ:CN:2008}. For instance, nano devices are unable to operate at microwave bands due to their small size. They will require molecular communication and Terahertz (THz) band for operation \cite{AKYILDIZ:CN:2008}. Additionally, in IoT devices, due to small energy sources, the computational processing capability is limited.  Therefore, it is necessary to meet the requirements for new protocols of nano devices at all layers of protocol stack. Operating in the THz band (0.1 - 10 THz) is a promising solution at the physical layer (PL) \cite{lemic:survey:2019} which makes the antenna size very small and thus suitable for exchanging information between nano devices. Potential applications of nanoscale communication in THz band include environmental monitoring, precision agriculture, smart health care and to name a few\cite{Mahboob:arXiv:2019}. 
\begin{figure*}[!htb]
\begin{center}
	\includegraphics[width=5in]{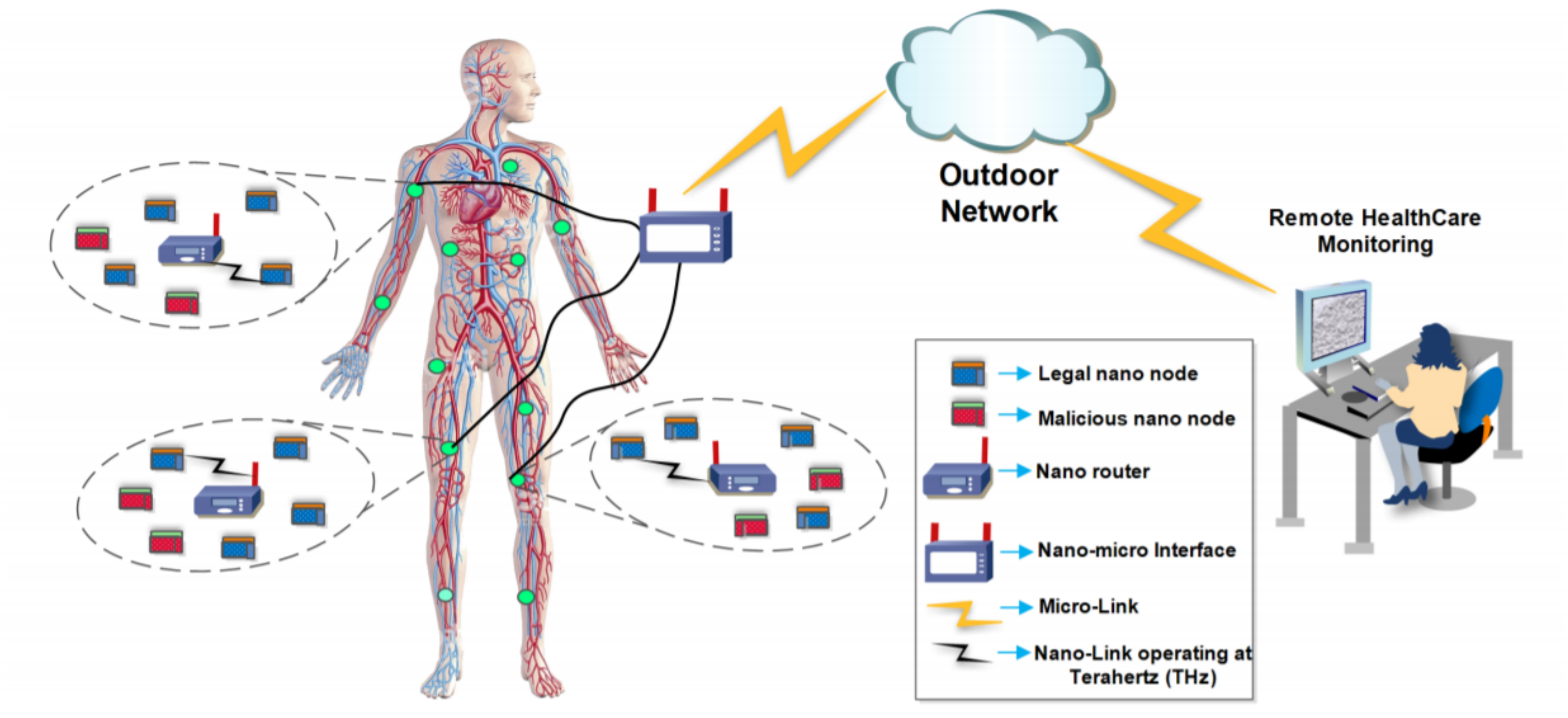} 
\caption{An envisioned  future nano-scale healthcare system with possible malicious nodes.}
\label{fig:app}
\end{center}
\end{figure*}

Like other communication networks the nanoscale communication networks are also prone to a wide range of active and passive attacks by adversaries\cite{Ma:Nature:2018}. Some of the common attacks include eavesdropping, impersonation, Denial of Services (DoS) etc. Here we investigate an impersonation attack in nanoscale communication networks. Figure \ref{fig:app} shows an illustration of a scenario of impersonation attack on a smart healthcare system. The nano nodes are deployed inside or on the body of a person for diseases diagnostics or to remotely monitor their health parameters. These nano devices are connected to a nano router which communicates the data to an outdoor network via a nano to micro interface. Assuming, an enemy of the person, secretly deployed its own nano nodes nearby with the aim of impersonating person's legal nodes to report false measurements to the remote monitoring system, an incorrect response through nano machines or nearby doctors could result in devastating consequences. Therefore, we need an authentication mechanism at the nano router to allow data transmission i.e. health measurements from legal nodes only, blocking all malicious nodes.

 In traditional communication systems, the countermeasures for such attacks were made at the higher layer using cryptography. Despite the wide work in the field of cryptography, the mechanism can be compromised because of its solely dependency on the predefined shared secret among the legal users. With recent advances in quantum computing, the traditional encryption become vulnerable to be easily decoded and existing crypto-based measures are not quantum secure unless the size of secret key increases to impractical length \cite{gidney:arXiv:2019}. In this regard, physical layer (PL) security finds itself a promising mechanism in future communication systems. PL security exploits the random nature of physical medium/layer for security purposes \cite{Poor:arXiv:2020}.\\
 Authentication is one of the pillars required for the security of any communication system. PL authentication is a systematic procedure that uses PL's features to provide authentication. In conventional systems, asymmetric key encryption (AKE) is typically used in the authentication phase which is the realm of public key encryption (a crypto based approach). Such schemes are quantum insecure and incur overhead or high computations which not only increase the size of the device but also consume high power. The devices fabricated for nanoscale communication are energy constrained as they comprise a small source of energy (battery). PL authentication has a low overhead (simple procedure which typically includes feature estimation and testing) and is almost impossible to clone unless the devices lie on each other. Various fingerprints including RSS \cite{Yang:TPDS:2013}, CIR \cite{waqas:UCET:2019} \cite{Aman:VTC:2017}, CFR \cite{Xiao:TWC:2008} \cite{Baracca:TWC:2012}, carrier frequency offset \cite{Hou:ICC:2012} \cite{Hou2:ICC:2012}, I/Q imbalance \cite{Hao:ICC:2014} are reported for PL authentication in  conventional communication systems.\\
 Regarding security of systems operating in THz band, we found two works\cite{Ma:Nature:2018}\cite{Mahboob:Access:2017} in the literature.  The experimental work of \textit{Jianjun et al.} \cite{Ma:Nature:2018} for the first time rejected the claim about security in THz band. The claim was that the inherit narrow beamwidth of THz link makes it secure and thus impossible for a malicious node to accomplish an eavesdropping attack. A similar claim is also made for beyond the THz band  \cite{Liang:NatureSR:2015}. The authors in \cite{Ma:Nature:2018} their experiments used reflectors of different shapes between THz transmitter and receiver. Then with the help of secrecy capacity and blockage as performance metrics, they clearly demonstrated that eavesdropping attack in THz can easily be done.The work considered an eavesdropping attack in a system operating in THz band which is different than the attack we consider in this work.
In our previous work \cite{Mahboob:Access:2017}, we studied PL authentication for an in-vivo nanoscale communication system whereby we utilized the path loss as the device fingerprint for three nodes system (i.e. Alice, Eve and Bob).  Our previous work \cite{Mahboob:Access:2017} was limited to three nodes system only. In this manuscript, we extend our previous work by studying the authentication of a generic system, comprising multiple legal and malicious nodes, operating in the THz band. We exploit the high-resolution transmission molecular absorption (HITRAN)\cite{Hitran:JQSRT:2017} data base for computing the path loss. We perform authentication by hypothesis testing. We refine the output of hypothesis testing via the hidden Markov model (HMM) viterbi algorithm. We also perform transmitter identification via the maximum likelihood and Gaussian mixture model (GMM) expectation maximization algorithm.

{\bf Outline.} The rest of this paper is organized as follows. Section II provides system model. Section III discusses authentication via two-step hypothesis testing. Section IV presents hidden markov model to refine the output of hypothesis testing. Section V provides transmitter identification schemes. Section VI presents simulation results with discussions and Section VII concludes the paper.
\section{System Model}
\begin{figure}[!htb]
\begin{center}
	\includegraphics[width=3in]{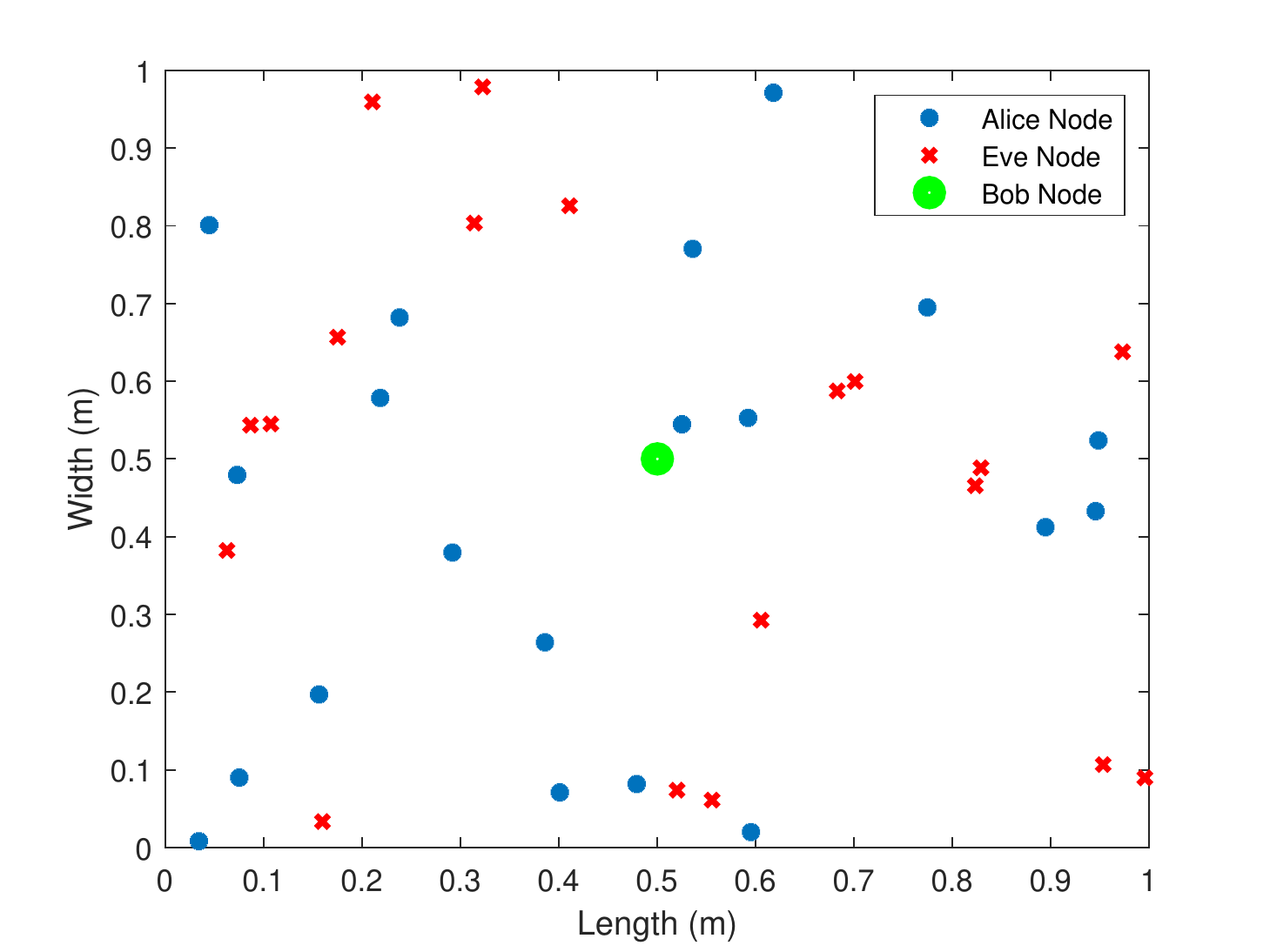} 
\caption{The system model: Bob is placed at origin. Alice and Eve nodes' locations are modelled as uniformly distributed random variables. In this case, $M=10$, $N=10$.}
\label{fig:sys-model}
\end{center}
\end{figure}

For the purpose of simulation, we considered a square 2D map/layout of size ($1$ m $\times 1$ m) where $M+N$ nano transmission (Tx) nodes, $M$ Alice (Legal) nodes $\{A_i\}_{i=1}^M$ and $N$ Eve (Malicous) nodes $\{E_j\}_{j=1}^N$, deployed according to the uniform distribution model, whilst a nano router/receiver node, Bob, is placed at origin as shown in Fig. \ref{fig:sys-model}. We assume that the Tx nodes transmits with a fixed/pre-specified transmit power so that the path loss can be computed by Bob.

The path loss is given as \cite{Josep:ICC:2010},\cite{Josep:TWC:2011}:

\begin{align}
\label{eq:THzPL}
L(f,d) [dB] = L_a(f,d)[dB] + L_s(f,d)[dB]
\end{align}

where $f$ is the frequency, $d$ is the distance, $L_a(f,d)[dB]$ is the absorption loss, and $L_s(f,d)[dB]$ is the spreading loss. 
The spreading loss is given as
\begin{equation}
L_s(f,d)[dB] = 20 \log_{10} (\frac{4 \pi f d}{c}),
\end{equation}
where $c$ is the speed of light. The absorption loss is given as:
\begin{equation}
L_a(f,d) = \frac{1}{\tau (f,d)}
\end{equation}
where $\tau$ represents the transmittance of the signal and is given by Beer-Lambart law:
\begin{eqnarray}
\tau(f,d) = e^{-k(f)d}
\end{eqnarray}
where $k$ is the medium absorption coefficient, given as:
\begin{equation}
k(f) = \sum_{i,g} k_{i,g}(f)
\end{equation}
where 
\begin{equation}
k_{i,g}(f) = \frac{p}{p_0}\frac{T_{0}}{T} Q_{i,g} \sigma_{i,g}(f) 
\end{equation}
where $i$ is the isotopologue (molecule that differs in isotropic composition) and $g$ is gas, $p_0(T_0)$ are standard pressure (Temperature), $\sigma_{i,g}(f)$ is the absorption cross-section and $Q_{i,g}$ is the molecular density given by
\begin{equation}
Q_{i,g} = \frac{n}{V} q_{i,g} N_A = \frac{p}{RT} q_{i,g} N_A
\end{equation}
where $R$ is gas constant, $N_A$ is the Avogadro constant and $q_{i,g}$ is the mixing ration for $i$ of $g$. The absorption cross-section can be expressed as
\begin{equation}
\sigma_{i,g} = S_{i,g} G_{i,g}(f)
\end{equation}
where, the line intensity $S_{i,g}$ and line shape $G_{i,g}(f)$ parameters can be computed using data from HITRAN database \cite{Hitran:JQSRT:2017}

\section{Authentication via Two-Step Hypothesis Testing}
We assume that the shared channel is time-slotted, whilst the transmit nodes perform channel sensing before transmitting, hence there are no collisions. Without the loss of generality, it can be assumed that $A_i$ is the legitimate user for the slot $k$, but if $A_i$ doesn't transmit during this time-slot, $E_j$ could transmit to Bob pretending to be an Alice node. Therefore, Bob needs to authenticate each message received on the shared channel and verify the transmitter identity (if no impersonation has been declared) in a systematic manner. 

Assuming that the noisy measurement $z(k)=L+n(k)$ has been obtained at time $k$ (for instance, by using the pulse based method as discussed in \cite{Josep:ICC:2012}), where $n(k)\sim N(0,\sigma^2)$ and $L$ is the path loss. Furthermore, in line with previous studies\cite{Mahboob:Access:2017},\cite{Mahboob:Globecom:2014}, we assume that Bob has already learnt the ground truth via prior training on a secure channel. The ground truth vector can be denoted by $\mathbf{l}=\{L_1,...,L_M\}^T$. The maximum likelihood (ML) hypothesis test can be explained by the following equation:  
\begin{equation} 
\label{eq:ML-pl}
(T^*,i^*) = \underset{i}{\min} \quad |z-L_i|
\end{equation}
Next, the binary hypothesis test works as follows: 
\begin{equation}
	\label{eq:H0H1}
	 \begin{cases} H_0 (\text{no impersonation}): & T^*=\underset{i}{\min} |z(k)-L_i| < \epsilon \\ 
                   H_1 (\text{impersonation}): & T^*=\underset{i}{\min} |z(k)-L_i| > \epsilon \end{cases}
\end{equation}
Equivalently, we have:
\begin{align} 
\label{eq:bht}
T^* \gtrless_{H_0}^{H_1} {\epsilon}
\end{align}
where $\epsilon$ is a small threshold - a design parameter. This work follows the Neyman-Pearson theorem \cite{Yan:IT:2001} which states that, for a pre-specified $P_{fa}$, $\epsilon$ can be chosen such that $P_{md}$ is minimized.

The error probabilities for the above hypothesis tests are:
\begin{equation}
\begin{split}
P_{fa} &= P(H_1|H_0)= \sum_{i=1}^M P(T^*>\epsilon|A_i)\pi (i) \\
&=\sum_{i=1}^M 2Q(\frac{\epsilon}{\sigma})\pi (i)=2Q(\frac{\epsilon}{\sigma})\sum_{i=1}^M \pi (i)=2Q(\frac{\epsilon}{\sigma})
\end{split}
\end{equation}
where $Q(x)=\frac{1}{\sqrt2\pi}\int_x^\infty e^{\frac{-t^2}{2}}dt$ is the complementary cumulative distribution function (CCDF) of a standard normal distribution, and $\pi (i)$ is the prior probability of $A_i$. Thus, the threshold could be computed as follows:

\begin{equation}
\label{eq:epsi}
\epsilon = \sigma Q^{-1}(\frac{P_{fa}}{2})
\end{equation}

Then, $P_{md}$ is given as:
\begin{equation}
\begin{split}
&{P}_{md}  = P(H_0|H_1) = P(T^*<\epsilon|H_1) \\
& =\sum_{j=1}^{N} \sum_{i=1}^{M} \bigg[ Q(\frac{L_i - L_j - \epsilon}{\sigma}) - Q(\frac{L_i - L_j + \epsilon}{\sigma}) \bigg] \pi (j)
\end{split}
\end{equation}
where $\pi (j)=\sum_{i=1}^M \alpha_{ij} \pi (i)$ is the prior probability of $E_j$. $0<\alpha_{ij}<1$ is the fraction of slots which are originally dedicated to $A_i$ but are found idle and thus utilized by $E_j$.

Since $P_{md}$ is an R.V., the expected value $\bar{P}_{md}:=\mathbb{E}(P_{md})$ is as follows:
\begin{equation}
\begin{split}
&\bar{P}_{md}  =\sum_{j=1}^N \frac{1}{\Delta}\pi (j). \\
& \bigg( \int_{L_{min}}^{L_{max}} \sum_{i=1}^M Q(\frac{L_i - L_j^{(E)} - \epsilon}{\sigma}) - Q(\frac{L_i - L_j^{(E)} + \epsilon}{\sigma}) dL_j^{(E)} \bigg) \\
& =\sum_{j=1}^N \frac{1}{\Delta}\pi (j). \\
& \bigg( \int_{L_{min}}^{L_{max}} \sum_{i=1}^M Q(\frac{L_i - L^{(E)} - \epsilon}{\sigma}) - Q(\frac{L_i - L^{(E)} + \epsilon}{\sigma}) dL^{(E)} \bigg)
\end{split}
\end{equation}
where we have assumed that the unknown path loss $L_j\sim U(L_{min},L_{max})$ $\forall j$, and $\Delta=L_{max}-L_{min}$. 

%\textsc{ Average KL Divergence.} As KL Divergence (KLD) is a measure of distance or distinction between two distributions and JSD is a measure of distance among multiple distributions. But here, the question is we have multiple legal and multiple illegal distributions so both measures are out of context. So we choose a new measure which is Average KLD (AKLD) and from Alices to Eves is given as
%\begin{align}
%{AKLD_{AE}}= \log_{10}(\frac{\sigma_A}{\sigma_E})+\frac{\sigma_A^2+(\mu_A-\mu_E)^2}{2 \sigma_E^2},
%\end{align}
%where
%\begin{align}
%&\mu_A=\frac{1}{T}\sum_i L_i, \ \ \mu_E=\frac{1}{T}\sum_j L_j, \nonumber\\ &\sigma_A^2=\sum_i (L_i-\mu_A)^2+T\sigma^2, \ \ \sigma_E^2=\sum_j (L_j-\mu_E)^2+T\sigma^2. \nonumber
%\end{align}
%with $T=M+N$  

\section{Hidden Markov Model based Approach}
At a given time instant $k$, the system is in one of the two states with the state-space: $\mathcal{S}=\{s_0,s_1\}$. The states $s_0$ and $s_1$ imply that there is no impersonation, impersonation respectively at time $k$. However, the true state of the system is hidden; therefore, what we observe through the hypothesis test is another observable Markov chain. The connection between the true/hidden state and the observable state is given by the emission probability matrix: 
\begin{equation}
\mathbf{R} = 
\begin{bmatrix}
  r_{0,0} & r_{0,1} \\
  r_{1,0} & r_{1,1}
 \end{bmatrix}
\end{equation}
where $r_{i,j}=Pr(x[k]=i|s[k]=j)$, $i,j \in \{ 0,1 \}$. The off-diagonal elements in the $i$-th row of $\mathbf{R}$ represents the errors made by the ML test, i.e., deciding the state as $s[k]=j$, $j \in \{0,1\}\setminus{i}$ while the system was actually in state $s[k]=i$.

The transition from state $i$ to state $j$ occurs after a fixed interval of $T=t_k-t_{k-1}$ seconds where $1/T$ is the measurement rate. Assuming that the system was in state $s_0$ at time $k=0$, i.e. $\mathbf{x}[0]=[1, 0]^T$ and we are in time $k-1$ and we want to predict the probability vector $\mathbf{x}[k]$ at time $k$ and the system is in state $s_i$, $i \in \{0,1\}$. To this end, we have the following transition probability matrix:
\begin{equation}
\mathbf{P} = 
\begin{bmatrix}
  p_{0,0} & p_{0,1} \\
  p_{1,0} & p_{1,1}
 \end{bmatrix}
\end{equation}
where $p_{i,j}=P(x[k]=j|x[k-1]=i)$, $i,j \in \{ 0,1 \}$.
Then, we have the following relation: $\mathbf{x}[k]=\mathbf{P}^k\mathbf{x}[0]$. Alternatively, we can write: $\mathbf{x}[k]=\mathbf{P}\mathbf{x}[k-1]$.

\subsection{ML Estimation of a Hidden Markov Sequence using Viterbi Algorithm}
Viterbi algorithm is used for ML sequence estimation (MLSE) of $\{s[k]\}_{k=1}^K$, given $\{x[k]\}_{k=1}^K$ as:
\begin{equation}
\{s[k]\} = \arg \max_{\{s^{'}[k]\}} p(x[k]|s^{'}[k])
\end{equation}

\section{Transmitter Identification} 
The probability of misclassification error resulting from Eq. \ref{eq:ML-pl} is given as: 
\begin{equation}
P_{mc} = \sum_{i=1}^M P_{mc|i}.\pi(i)
\end{equation}
where $P_{mc|i}=P(\text{Bob decides }A_j|A_i\text{ was the sender})$. For the hypothesis test of (\ref{eq:bht}), $P_{mc|i}$ is given as:
\begin{equation}
\label{eq:pmc}
P_{mc|i}=1-\bigg( Q(\frac{\tilde{L}_{l,i}-\tilde{L}_i}{\sigma}) - Q(\frac{\tilde{L}_{u,i}-\tilde{L}_i}{\sigma}) \bigg)
\end{equation}
where $\tilde{L}_{l,i}=\frac{\tilde{L}_{i-1}+\tilde{L}_i}{2}$, $\tilde{L}_{u,i}=\frac{\tilde{L}_{i}+\tilde{L}_{i+1}}{2}$. Additionally, $\mathbf{\tilde{l}}=\{\tilde{L}_{1},...,\tilde{L}_{M}\}=\text{sort}(\mathbf{l})$ where sort operation (.) sorts a vector in an increasing order. For the boundary cases, e.g., $i=1, i=M$, $\tilde{L}_{l,1}=L_{min}$, $\tilde{L}_{l,M}=L_{max}$ respectively.

\subsection{Transmitter Identification using Gaussian Mixture Modelling}
The GMM was consisted of $Q=M+N$ component densities where only the $Q=M$ densities could be trained. The $3Q$ GMM parameters was learnt by running the Expectation-Maximization (EM) algorithm on the training data. GMM, in its standard form, is perfectly suited for transmitter identification.
Under GMM, the probability density function (pdf) of the (observed) mixture random variable $X$ is the convex/weighted sum of the component pdfs:
\begin{equation}
f_X(x) = \sum_{q=1}^Q \pi_q \phi_q(x)
\end{equation}
where each $\phi_q(x)$ is a Gaussian pdf which satisfies: $\phi_q(x)\geq 0$, $\int_{x\in \mathbb{R}} \phi_q(x)dx=1$. The weights/priors satisfy: $\pi_q(x)\geq 0$, $\sum_{q=1}^Q \pi_q=1$.

The GMM has $3Q$ unknown parameters which were learnt by applying the iterative Expectation-Maximization algorithm on training data $\{x_m\}_{m=1}^{M}$.
The posterior probability for each point $x_m$ in the training data (i.e., the likelihood of $x_m$ belonging to component $q$ of the mixture) was computed as follows ($j$ is the iteration number):
\begin{equation}
\label{eq:em1}
p_{m,q}^{(j)} = \frac{ \pi_q^{(j)} \phi_q(x_m, \mu_q^{(j)}, \Sigma_q^{(j)}) } {\sum_{\hat{q}=1}^{Q} \pi_{\hat{q}}^{(j)} \phi(x_m, \mu_{\hat{q}}^{(j)}, \Sigma_{\hat{q}}^{(j)})}
\end{equation}
The $Q$ number of priors were updated as follows:
\begin{equation}
\pi_q^{(j+1)} = \frac{1}{M} \sum_{m=1}^{M} p_{m,q}^{(j)}
\end{equation}
The $Q$ number of means were updated as follows:
\begin{equation}
\mu_q^{(j+1)} = \frac{ \sum_{m=1}^{M} p_{m,q}^{(j)} x_m } { \sum_{m=1}^{M} p_{m,q}^{(j)} }
\end{equation}
The $Q$ number of (co-)variances were updated as follows:
\begin{equation}
\label{eq:em4}
\Sigma_q^{(j+1)} = \frac{ \sum_{m=1}^{M} p_{m,q}^{(j)} (x_m-\mu_q^{(j)})(x_m-\mu_q^{(j)})^T } { \sum_{m=1}^{M} p_{m,q}^{(j)} }
\end{equation}
The iterative EM algorithm monotonically increased the objective (likelihood) function value, and converged when the increase in likelihood function value between two successive iterations became less than the threshold $\epsilon$.

\section{Simulation Results}
%\begin{figure}[ht]
%\begin{center}
%	\includegraphics[width=3.8in]{Figures/kfvsf_H2o.eps} 
%\caption{The path loss model (reproduced from \cite{Josep:TWC:2011}) }
%\label{fig:Kvsf}
%\end{center}
%\end{figure}
We kept $M = N = 10$, $\alpha_{ij}= 0.5 \ \forall j$, $f=1$ THz, $T=285$ k and $p=1$ atm. Both Alice and Eve nodes were deployed according to uniform distribution in $1m \times 1m$ area. Total $10^5$ random realisations (independent for Alice and Eve nodes) of nodes' deployment were taken and then errors were averaged over the realizations. 

\begin{figure}[htb!]
\begin{subfigure}{.5\textwidth}
  \centering
  \includegraphics[width=1\linewidth,height=60mm]{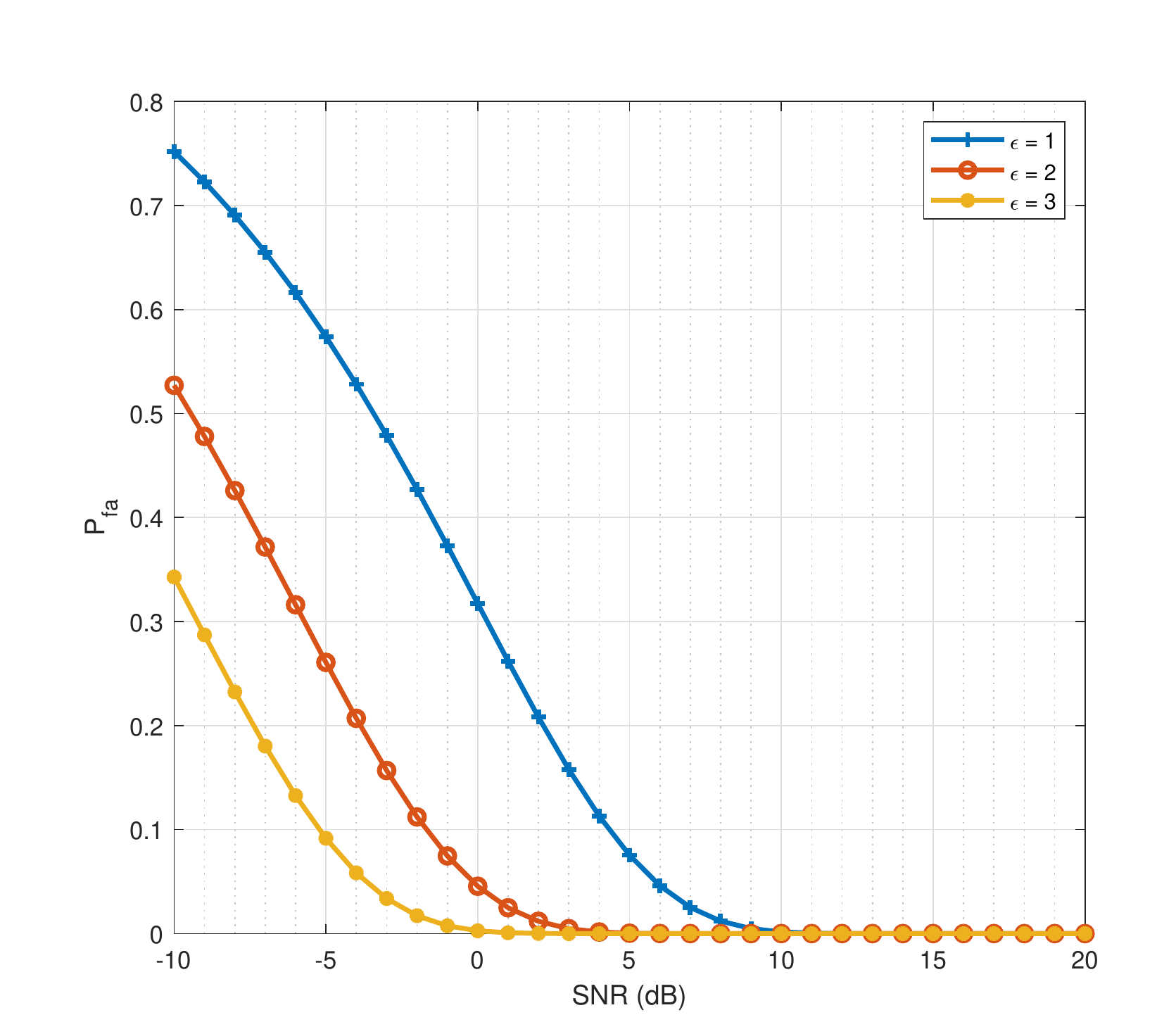}
  \caption{False alarm}
  \label{fig:sfig1}
\end{subfigure}%
\\
\begin{subfigure}{.5\textwidth}
  \centering
  \includegraphics[width=1\linewidth,height=60mm]{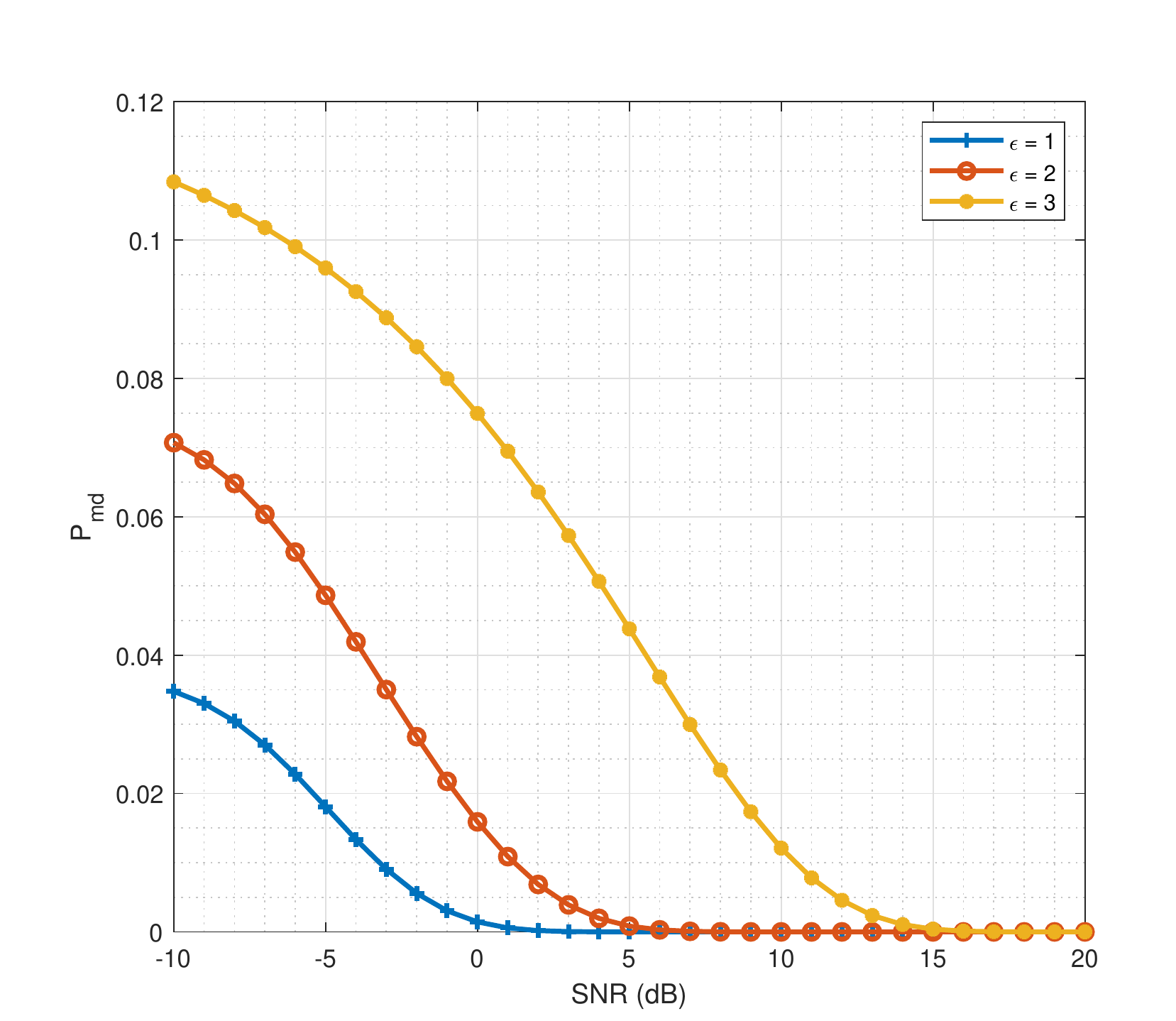}
  \caption{Missed detection}
  \label{fig:sfig2}
\end{subfigure}
\caption{The error probabilities against the SNR $\frac{1}{\sigma^2}$}
\label{fig:Pevssnr}
\end{figure}
$P_{fa}$ and $P_{md}$ are  two well-known probabilities resulting in hypothesis testing. The $P_{fa}$ was defined as the probability that any $i-$th Alice node can be considered as any of the Eve nodes. While the $P_{md}$ is the probability of the event that any $j-$th Eve node can be considered as any of the Alice nodes.  

Fig. \ref{fig:Pevssnr} represents the two probabilities against the SNR $=\frac{1}{\sigma^2}$ where the improvement in error probabilities with an increasing SNR can be seen clearly. The  designed parameter $\epsilon$ decreases ${P}_{md}$ but increases $P_{fa}$.

\begin{figure}[htb!]
\begin{subfigure}{.5\textwidth}
  \centering
  \includegraphics[width=1\linewidth,height=60mm]{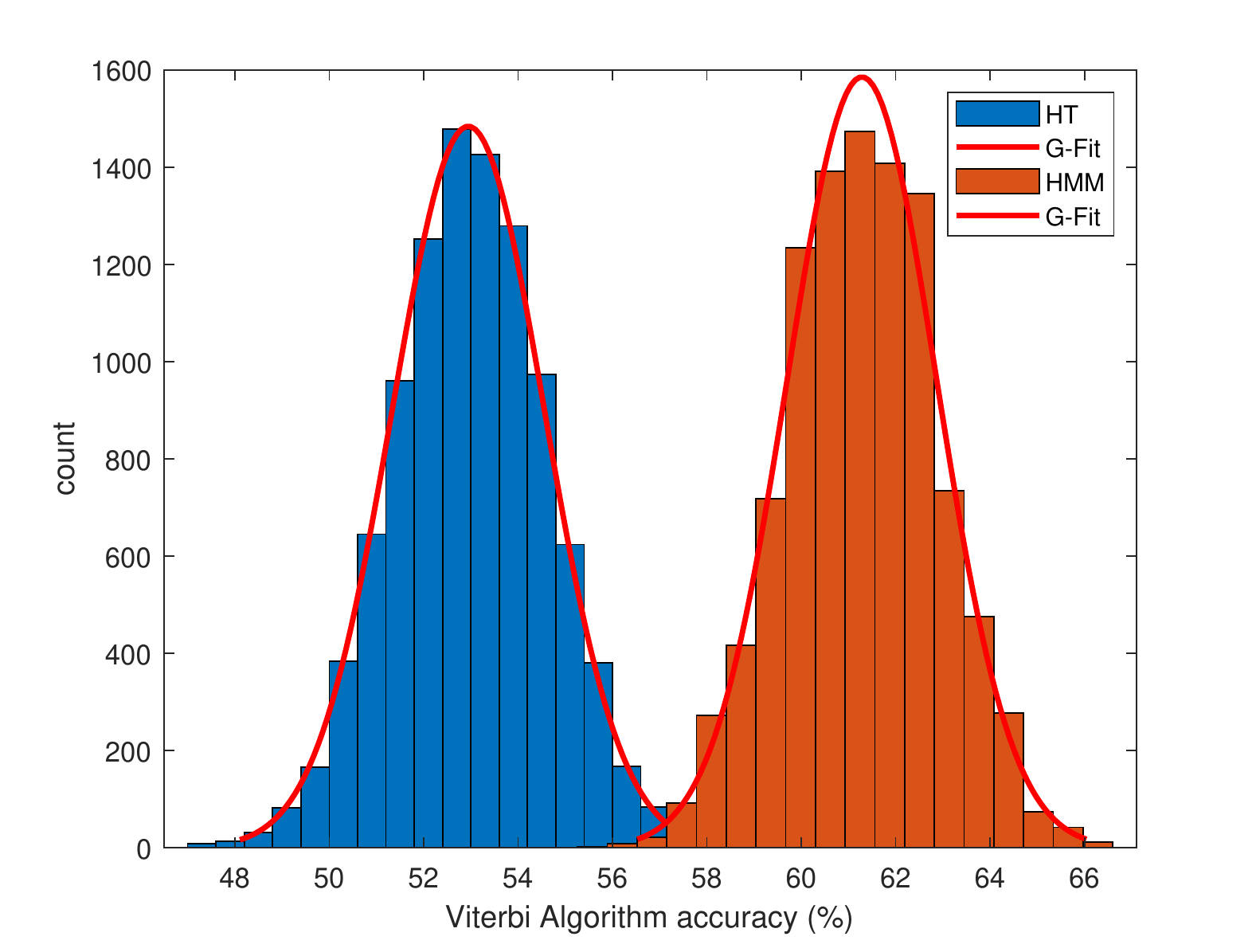}
  \caption{SNR $= -5$ dB}
  \label{fig:HMMfig1}
\end{subfigure}%
\\
\begin{subfigure}{.5\textwidth}
  \centering
  \includegraphics[width=1\linewidth,height=60mm]{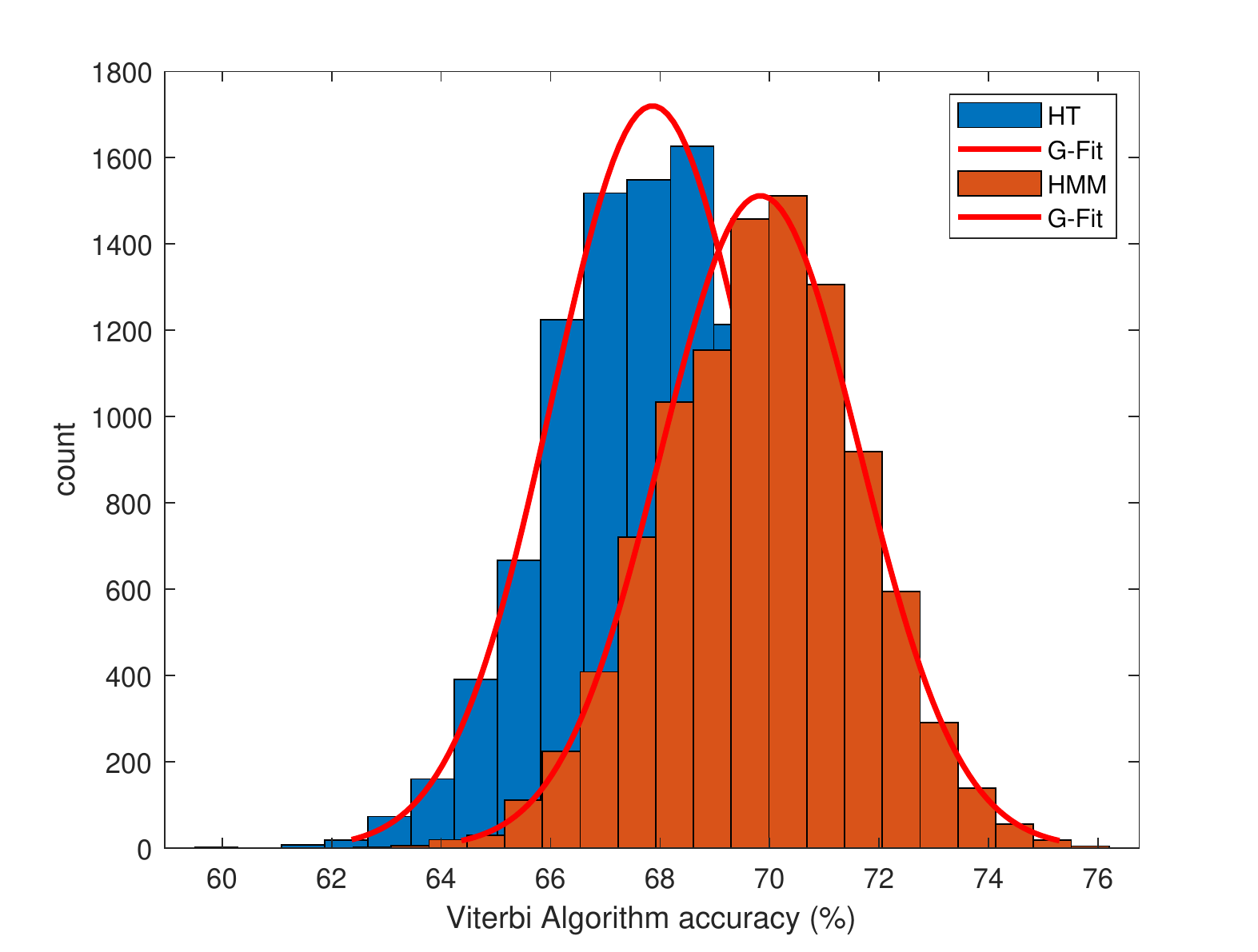}
  \caption{SNR $= 0$ dB }
  \label{fig:HMMfig2}
\end{subfigure}
\caption{Performance comparison of HT and HMM.}
\label{fig:HMM}
\end{figure}

%\begin{figure}[!htb]
%\centering
%  \begin{tabular}{@{}cc@{}}
%    \includegraphics[width=.5\textwidth]{Figures/HMM_5db.eps} 
%    \includegraphics[width=.5\textwidth]{Figures/HMM0dB.eps} 
%  \end{tabular}
%  \caption{Performance comparison of HT and HMM. $SNR = -5 dB$ is taken in left plot while $SNR = 0dB$ in right plot.}
%  \label{fig:HMM}
%\end{figure}

\begin{figure*}[htb!]
\begin{subfigure}{.5\textwidth}
  \centering
  \includegraphics[width=1\linewidth,height=60mm]{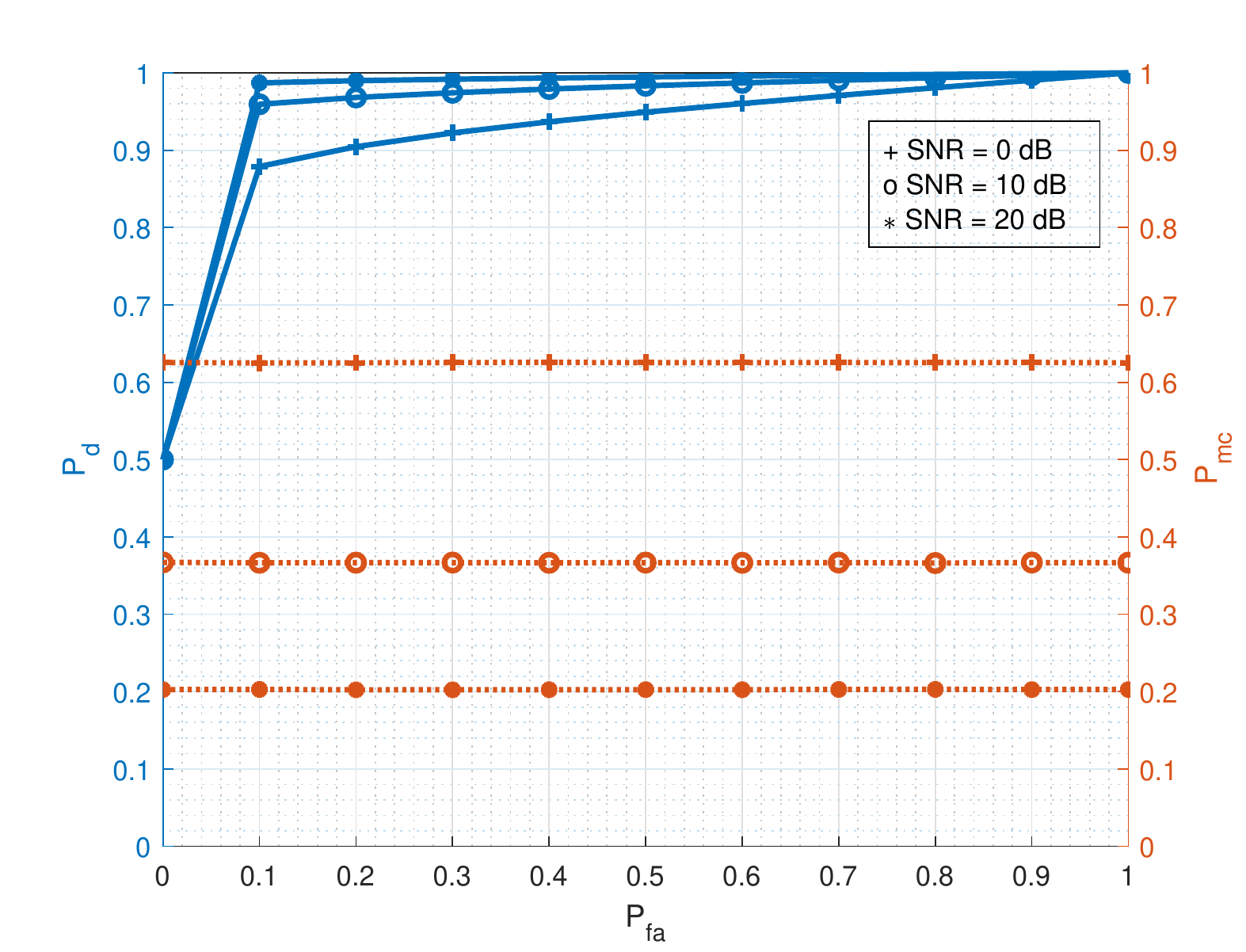}
  \caption{$M = N = 10$ $\alpha_{ij} = 0.5 \ \forall j$}
  \label{fig:ROCfig1}
\end{subfigure}%
\begin{subfigure}{.5\textwidth}
  \centering
  \includegraphics[width=1\linewidth,height=60mm]{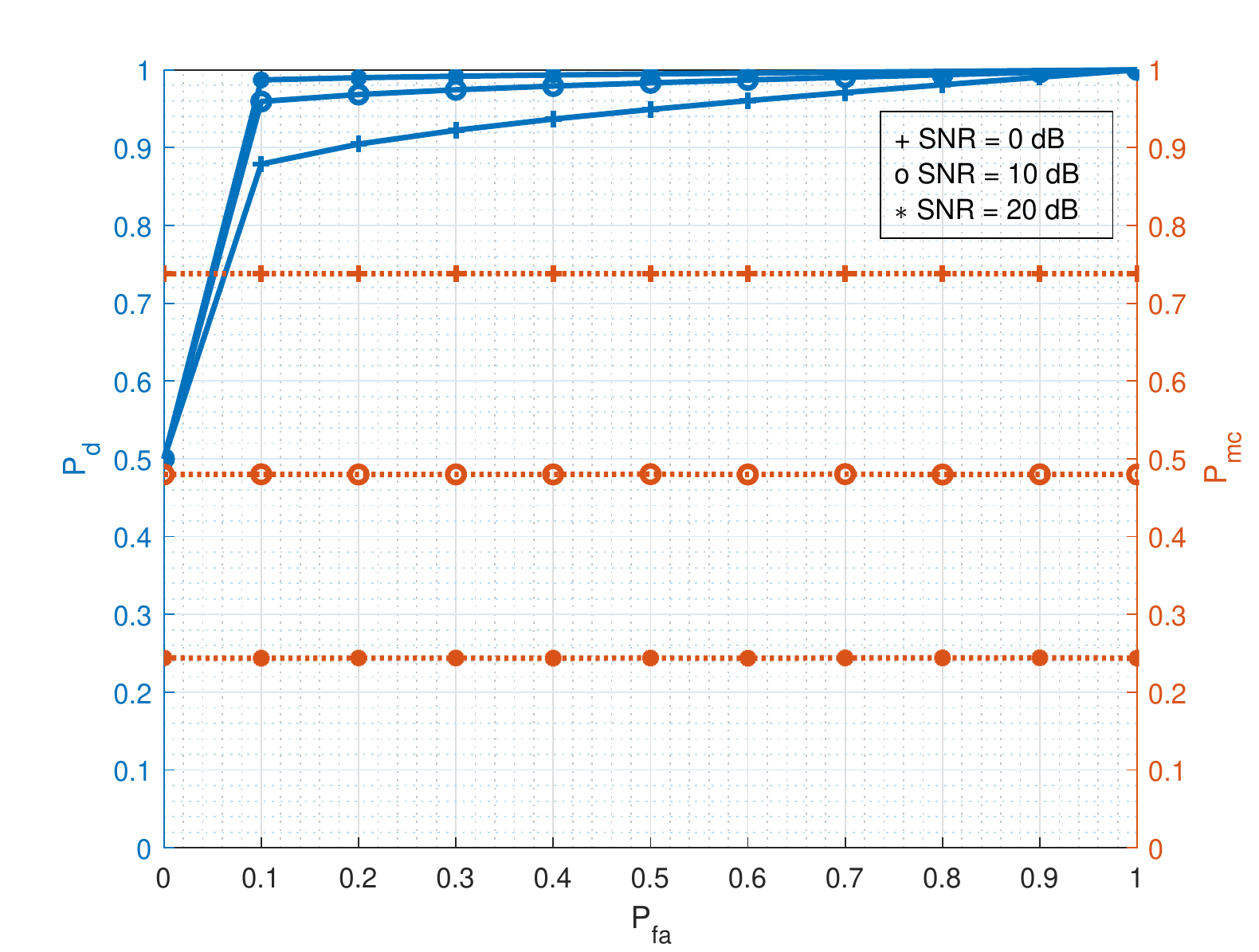}
  \caption{$M = N = 20$ $\alpha_{ij} = 0.5 \ \forall j$}
  \label{fig:ROCfig2}
\end{subfigure}
\begin{subfigure}{.5\textwidth}
  \centering
  \includegraphics[width=1\linewidth,height=60mm]{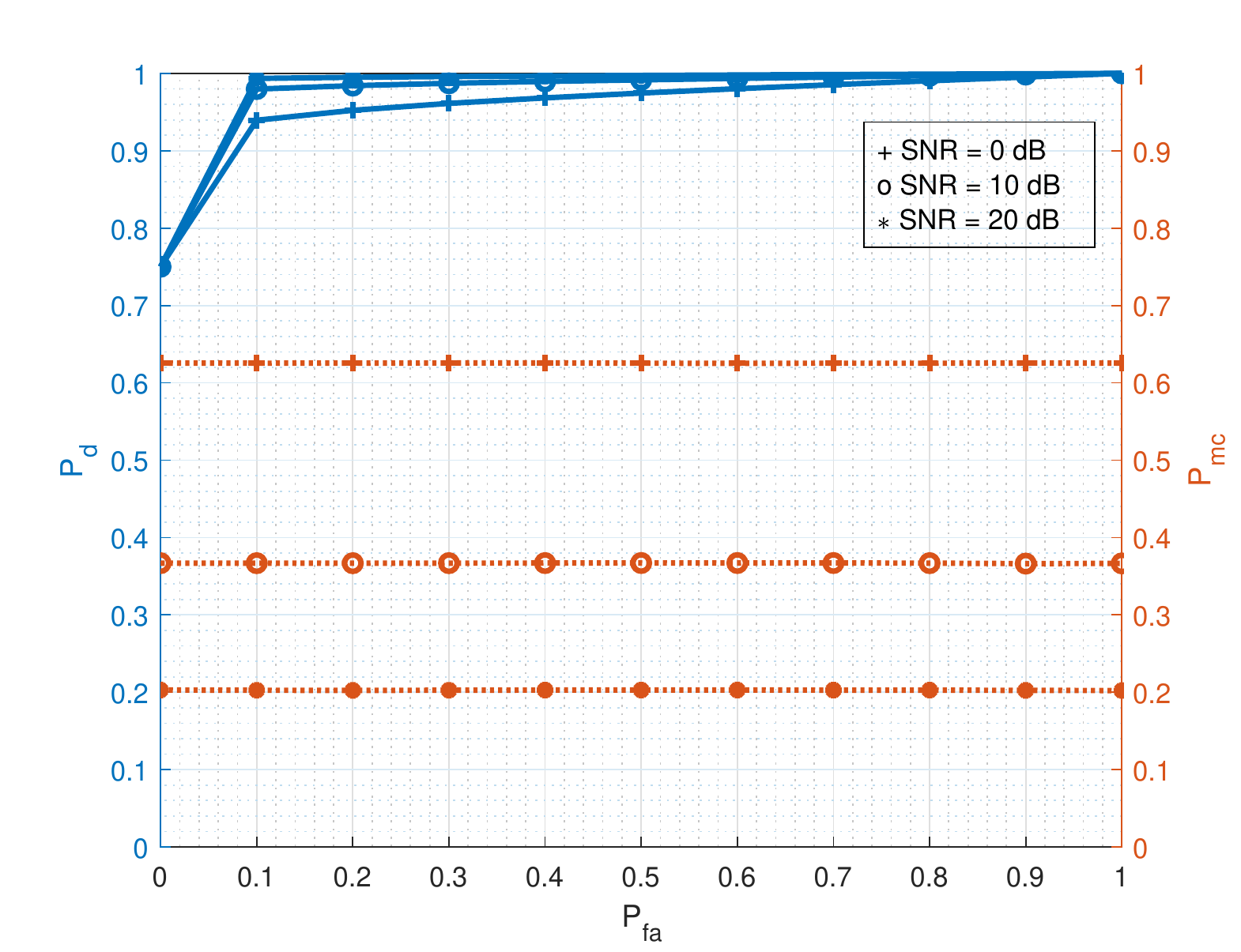}
  \caption{$M = N = 10$ $\alpha_{ij} = 0.25 \ \forall j$}
  \label{fig:ROCfig3}
\end{subfigure}%
\begin{subfigure}{.5\textwidth}
  \centering
  \includegraphics[width=1\linewidth,height=60mm]{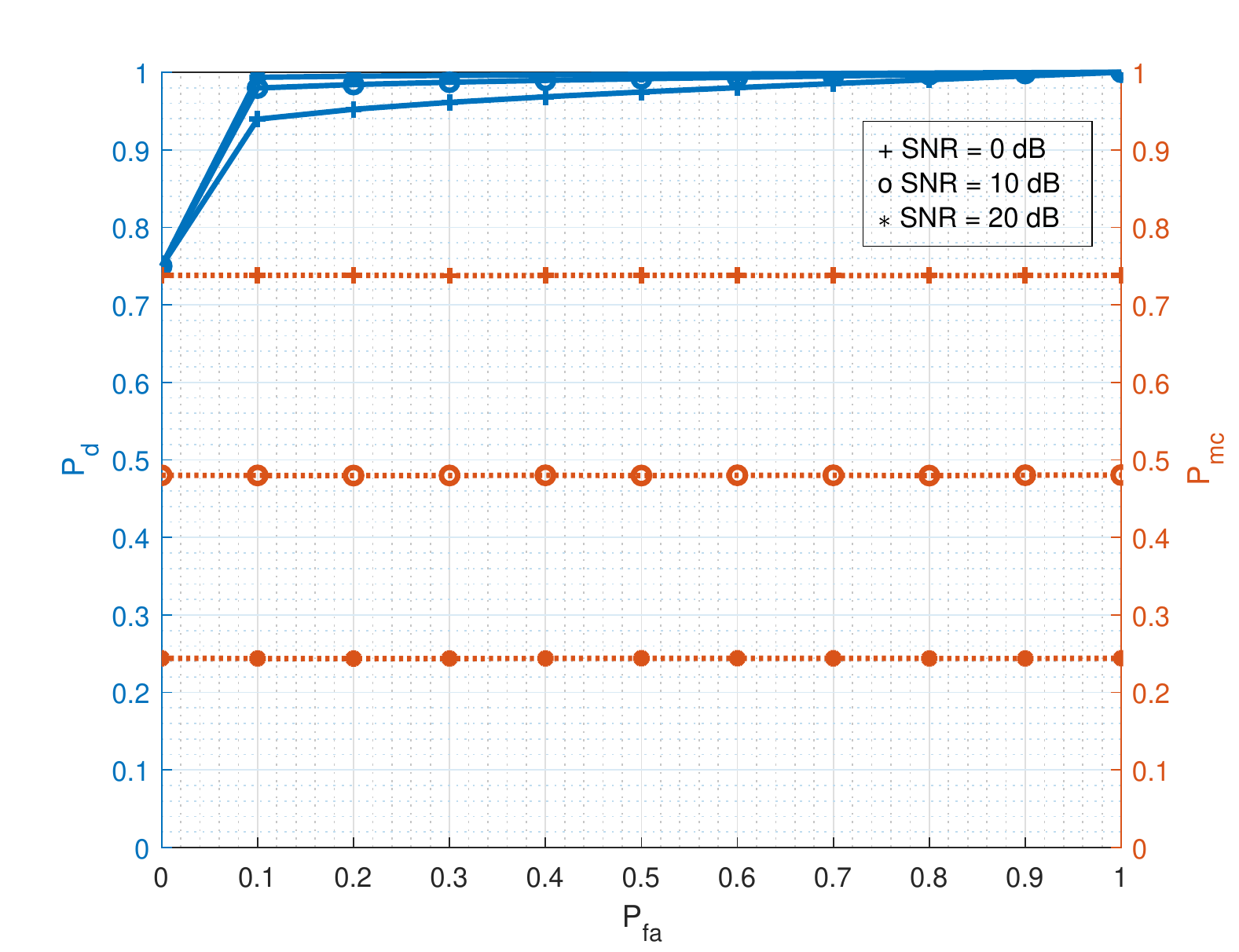}
  \caption{$M = N = 20$ $\alpha_{ij} = 0.25 \ \forall j$}
  \label{fig:ROCfig4}
\end{subfigure}
\caption{ROC curves}
\label{fig:ROC}
\end{figure*}

%\begin{figure}[!htb]
%\centering
%  \begin{tabular}{@{}cc@{}}
%    \includegraphics[width=.5\textwidth]{Figures/ROCm10a0.5.eps} &
%    \includegraphics[width=.5\textwidth]{Figures/ROCm20a0.5.eps} \\
%    \includegraphics[width=.5\textwidth]{Figures/ROCm10a0.25.eps} &
%    \includegraphics[width=.5\textwidth]{Figures/ROCm20a0.25.eps}
%  \end{tabular}
%  \caption{ROC. First row is for $\alpha_{ij} = 0.5 \forall j $ while $M$ and $N$ changes from $10$ to $20$ from left to right. Similar goes with second roo with $\alpha_{ij} = 0.25 \forall j$.}
%  \label{fig:ROC}
%\end{figure}
Fig. \ref{fig:HMM} shows the efficacy of HMM. At low SNR the performance of HMM was far better than HT and at high SNR HT was closed to HMM. The results were produced after monte carlo based simulation. The total number of transmissions were kept to $10^5$ (more specifically, $10^5$ binary states ($s_0$, $s_1$) were generated), $\epsilon = 1$, $\mathbf{P}=0.5I_{2\times 2}$, where $I$ is the identity matrix and $K=10^3$. The error resulting from HT and HMM methods were calculated as the number of times the predicted/estimated state was not equal to actual state divided by total transmissions. The accuracy was then computed accordingly. The entries of $\mathbf{R}$ were calculated according to $P_{fa}$ and $P_{md}$. 
Fig. \ref{fig:ROC} shows the Receiver Operating characteristic (ROC) curves for different configurations of node and transmissions from Eve nodes (i.e. $\alpha_{ij}$). Typically, the ROC contains two error probabilities ($P_d$ and $P_{fa}$), but due to multiple nodes in this study, we had three probabilities. For any $P_{fa}$ value the $P_{mc}$ was constant which was obvious from Eq.\ref{eq:pmc}. Increasing SNR not only improved $P_d$ but also improved $P_{mc}$ as well. The $P_{fa}$ was chosen as an independent variable and swept on the range $0$ to $1$. Using Eq. \ref{eq:epsi}, the threshold was calculated for a given SNR value. Further, $P_d=1-P_{md}$ (the detection probability) and $P_{mc}$ were computed as average after doing $10^5$ uniform realisations of nodes deployment. We observed that increasing the number of nodes does not affect the $P_d$ but $P_{mc}$ increases with an increase in number of Alice nodes ($M$). We further observed that the less the nodes (Alice nodes) remain idle during their allocated slot the more the $P_d$ we had. 

\begin{figure}[htb!]
\begin{subfigure}{.5\textwidth}
  \centering
  \includegraphics[width=1\linewidth,height=60mm]{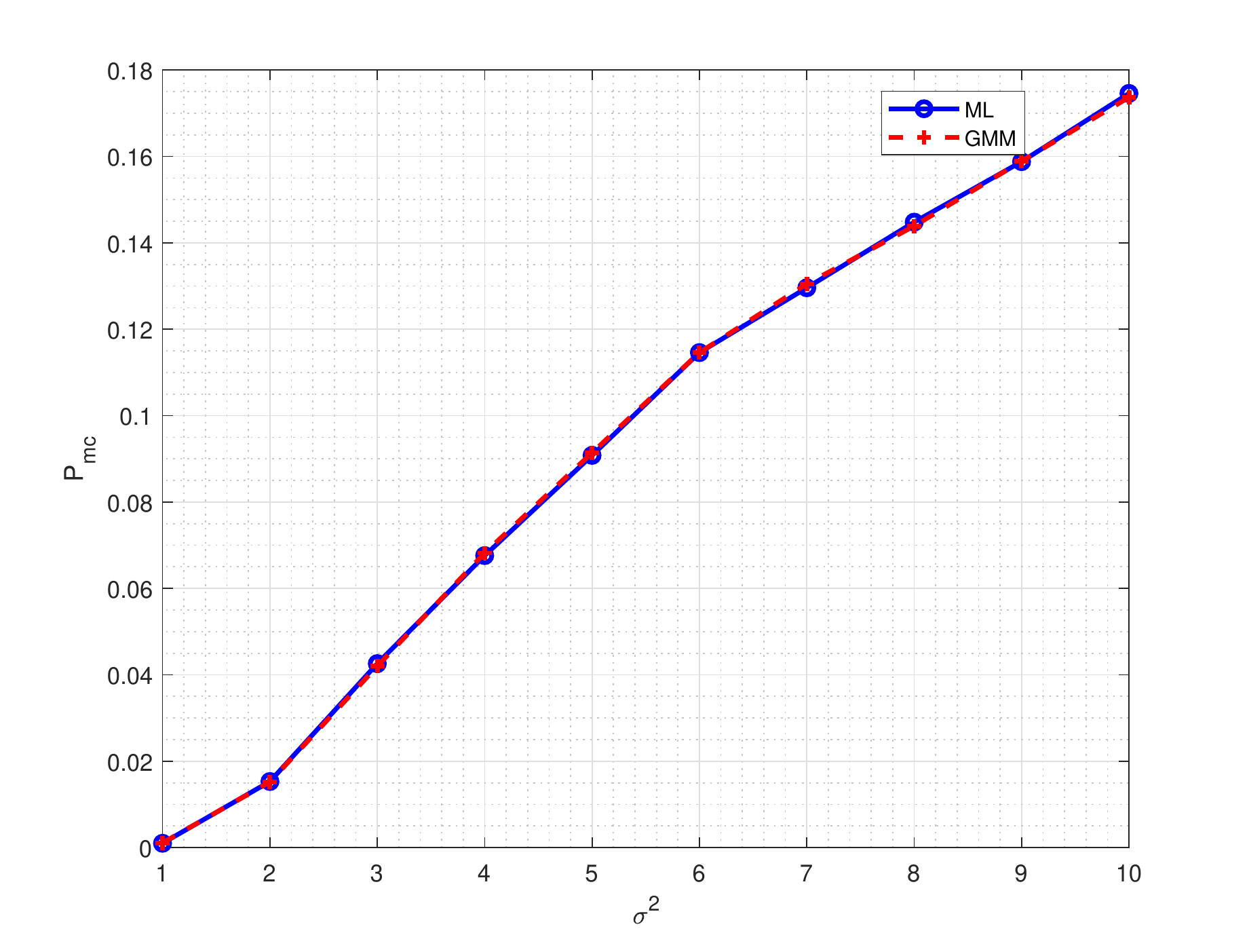}
  \caption{Noiseless ground truths}
  \label{fig:GMMfig1}
\end{subfigure}%
\\
\begin{subfigure}{.5\textwidth}
  \centering
  \includegraphics[width=1\linewidth,height=60mm]{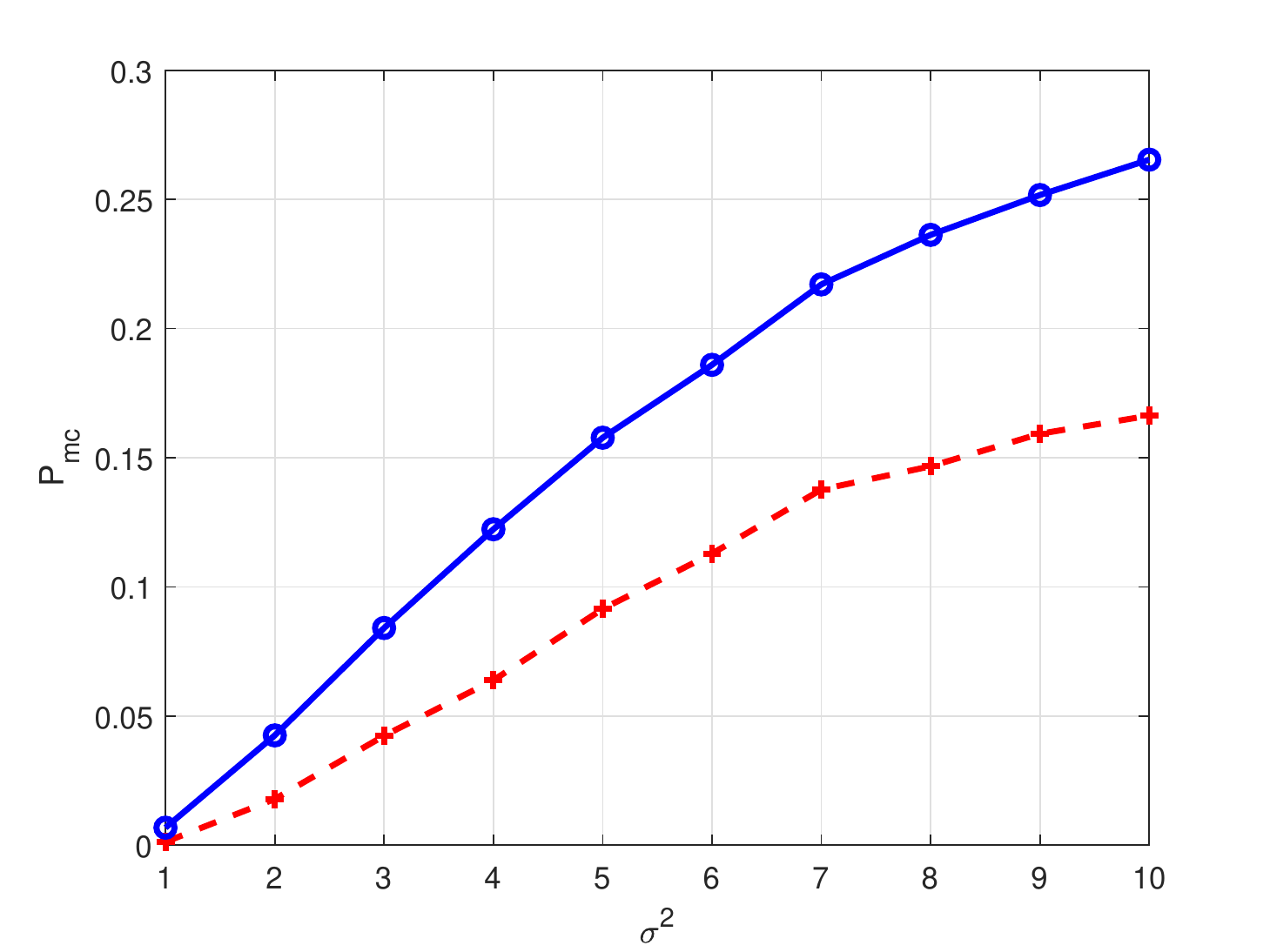}
  \caption{Nosiy ground truths}
  \label{fig:GMMfig2}
\end{subfigure}
\caption{$P_{mc}$ vs estimation error $\sigma^2$.}
\label{fig:GMM1}
\end{figure}

$P_{mc}$ is the probability of deciding $i-$th Alice node, as any Alice node without $i$. $P_{mc}$ becomes an important metric when dealing with multiple nodes' identification. Here, $P_{mc}$ resulted from both transmitter identification's algorithms (ML which is a bi-product of two-step HT based authentication and GMM).  
As GMM is a learning approach, it requires training data to learn its parameters. That is the reason that we only performed transmitter identification using GMM. We assumed no data was available for Eve nodes.

%\begin{figure}[!htb]
%  \begin{tabular}{@{}cc@{}}
%    \includegraphics[width=.5\textwidth]{Figures/GMM+ML.eps} &
%    \includegraphics[width=.5\textwidth]{Figures/GMM+ML_NoisyFP.eps} 
%  \end{tabular}
%  \caption{$P_{mc}$ vs estimation error $\sigma^2$. (a) left plot and (b) right plot.}
%  \label{fig:GMM1}
%\end{figure} 
Fig. \ref{fig:GMM1} (a) was generated by assuming actual ground truths (noiseless $(L_i \ \forall i)$) of Alice nodes available for performing ML based transmitter identification. The ML was implemented using Eq.\ref{eq:ML-pl} with having noiseless ground truths. Fig. \ref{fig:GMM1} (a) shows that the two approaches perform equally. To test the efficacy of the GMM approach, we performed another experiment and plotted the results in Fig. \ref{fig:GMM1} (b). This time we assumed that the ground truths of Alice nodes were noisy  $L_i +n \ \forall i$ (i.e when the ground truths were obtained on secure channel it also included noise or an error). This time the the ML based approach was implemented using Eq.\ref{eq:ML-pl} to include noisy ground truths. The GMM parameters were estimated on $10^4$ training data generated from the legal nodes and then tested on $10^5$. The error was calculated as number of times the estimated state was not equal to the actual value divided by the total transmissions for both approaches and for both cases. We observed from Fig. \ref{fig:GMM1} (b) that the overall performance of GMM was improved. The performance improved even further for lower SNR or higher $\sigma^2$. 
\section{Conclusion}
This paper provided an authentication mechanism using path loss as a fingerprint at the physical layer in nanoscale communication systems operating in terahertz band. The work's importance was advocated by illustrating envisioned smart healthcare application of nanoscale communication systems. The complex and quantum insecure crypto measures can be complemented using this approach which is simple and quantum secure (i.e. no encryption or shared secret key is involved). This was observed from ROC curves after doing monte carlo based simulation for nodes deployment under uniform distribution that with $20 \%$ false rate the detection probability is almost one when operating with SNR $= 10$ dB. A generic system was considered  comprising multiple legal and malicious nano nodes operating in Terahetrz band. Specifically, for simulation purpose, nodes were deployed in $1m \times 1m$ area under uniform distribution and air was considered as a medium among the nodes and path loss was calculated using HITRAN data base.   
\section*{Acknowledgements}
This work was made possible by NPRP grant number NPRP 10-1231-160071 from  the Qatar National Research Fund (a member of Qatar Foundation). The statements made herein are solely the responsibility of the authors.  Waqas Aman would also like to thank Higher Education Commission of Pakistan for providing him IRSIP scholarship to travel to University of Glasgow for his studies.

\footnotesize{
\bibliographystyle{IEEEtran}
\bibliography{main}
}

\vfill\break

\end{document}